\documentclass{article}
\usepackage{graphicx} 
\usepackage{xcolor}

\title{Generative AI for Social Impact}
\author{Lingkai Kong, Cheol Woo Kim, Davin Choo, Milind Tambe}
\date{December 2025}

\usepackage{url}
\usepackage{amsmath}

\begin{document}

\maketitle

\section*{Abstract}

AI for Social Impact (AI4SI) has achieved compelling results in public health, conservation, and security, yet scaling these successes remains difficult due to a persistent deployment bottleneck. We characterize this bottleneck through three coupled gaps: observational scarcity resulting from limited or unreliable data; policy synthesis challenges involving combinatorial decisions and nonstationarity; and the friction of human–AI alignment when incorporating tacit expert knowledge and dynamic constraints. We argue that Generative AI offers a unified pathway to bridge these gaps. LLM agents assist in human-AI alignment by translating natural-language guidance into executable objectives and constraints for downstream planners, while diffusion models generate realistic synthetic data and support uncertainty-aware modeling to improve policy robustness and transfer across deployments. Together, these tools enable scalable, adaptable, and human-aligned AI systems for resource optimization in high-stakes settings.

\section{Introduction}

\textbf{AI for Social Impact (AI4SI)} is a distinct and rapidly growing subdiscipline of artificial intelligence research dedicated to achieving measurable positive societal impact, particularly for vulnerable and under-resourced populations~\cite{shi2020artificial, perrault2020artificial}. This field leverages the predictive, optimization, and decision-making capabilities of AI to address complex global challenges in areas like public health~\cite{yadav2016using}, environmental sustainability~\cite{gomes2019computational}, disaster response~\cite{madaio2016firebird}, and poverty alleviation~\cite{jean2016combining}. The fundamental objective of AI4SI is often the design and deployment of systems that can optimize limited resources to achieve maximal human benefit.

Translating these objectives into practice requires addressing the central challenge of resource optimization under uncertainty. Over the past two decades, our work has focused on developing algorithmic frameworks to meet this challenge across diverse high-stakes domains, including public health (maternal and child care~\cite{verma2023}, HIV prevention~\cite{rice2021peer}), wildlife conservation~\cite{yang2014adaptive}, and public safety~\cite{pita2008deployed}. These deployments provide empirical evidence that AI-driven optimization can yield significant, verifiable improvements in real-world settings. For example, in large-scale mobile health programs in India, these strategies contributed to a 30\% reduction in program drop-out rates~\cite{verma2023}, while in national parks, they led to a five-fold increase in the detection of illegal snares~\cite{xu2020stay}.

Despite these notable successes in algorithmic design and localized deployment, a persistent and critical barrier stands in the way of achieving widespread, scalable AI4SI adoption: \textbf{the deployment bottleneck}. This bottleneck  manifests across three distinct but interconnected gaps: the observational scarcity gap (data acquisition), the policy synthesis gap (modeling and learning), and the human-AI alignment gap (real-world implementation). This paper asserts that the next generation of AI4SI scalability will be unlocked by the strategic application of \textbf{Generative AI} to overcome each facet of this bottleneck.

The rest of this paper is organized as follows. Section~2 reviews representative AI4SI deployments from our group and highlights how the deployment bottleneck arises in practice. Section~3 discusses how generative AI, via LLM agents and diffusion models, can bridge the three gaps and enable scalable, human-aligned deployment. Section~4 concludes with open directions toward trustworthy AI4SI at scale.

\section{Deployment Bottlenecks}

\subsection{The Observational Scarcity Gap (Data)}
A key challenge in many AI4SI domains is the observational scarcity, which makes it difficult to have the required data for learning and inference. We illustrate this challenge via our continued effort on network-based HIV prevention. 

\paragraph{YEH in Los Angeles: Hidden Social Networks.} Our team conducted one of the first large-scale applications of social network algorithms for public health, focusing on spreading HIV prevention information among \textbf{Youth Experiencing Homelessness (YEH)} in Los Angeles starting in 2013 \cite{rice2021peer}.
This pioneering field study demonstrated the use of \textbf{AI-guided interventions} to achieve significant reductions in HIV risk behaviors among 750 YEH compared to traditional methods.
However, this work provided a clear, early example of the \textbf{observational scarcity gap}.
The optimal intervention strategy depended entirely on the hidden, unobserved social network structure linking the youth.
Because this network was not known in advance, our influence maximization algorithm had to rely on a sampled fraction of the social network \cite{wilder2018maximizing,kamarthi2020influence}.

\paragraph{South Africa: Partially Observed Contact Networks.}
In continued work on HIV prevention in South Africa, our team confronted the observational scarcity gap in an even more direct form.
Although network-based testing is endorsed by the WHO as a key strategy \cite{world2024consolidated}, the underlying transmission/contact network is rarely observed in full in real deployments. Instead, teams act sequentially, observing infection statuses only as individuals are tested. Our Adaptive Frontier Exploration on Graphs (AFEG) framework~\cite{choo2025adaptive} formalizes this setting and yields principled decision rules \emph{given} a contact graph, but the core scarcity remains: AFEG assumes the graph is known upfront, whereas real deployments often face missing or unreliable network structure.

\subsection{The Policy Synthesis Gap (Learning and Modeling)}
Even with sufficient data, learning deployment-ready policies in AI4SI remains challenging. Decisions often depend on high-dimensional state information and \emph{combinatorial} action spaces, where naive policy parameterizations (e.g., assigning probability to every action) are computationally infeasible at scale. Moreover, policies learned purely from historical data can be brittle under environmental change and domain shift. We refer to this mismatch between deployment requirements and what standard learning pipelines can reliably deliver as the \textbf{policy synthesis gap}: learning policies that are simultaneously near-optimal, scalable, and resilient to nonstationarity.

\paragraph{Patrol Planning in Wildlife Conservation.}
Our work on the Protection Assistant for Wildlife Security (PAWS)~\cite{yang2014adaptive} supports ranger patrol planning in large protected areas~\cite{xu2020stay}. Here, patrol teams repeatedly select a subset of regions to visit each week (or month) based on poaching-risk forecasts. Even with a modest discretization, the action space is inherently combinatorial: selecting $5$ regions out of $50$ yields $\binom{50}{5}=2{,}118{,}760$ possible subsets, making explicit enumeration or full-distribution policies impractical. Beyond scale, deployed policies must remain effective as adversaries adapt and risk landscapes shift, so nonstationarity can quickly erode performance. Bridging this gap requires methods that (i) scale to large combinatorial decision spaces and (ii) adapt rapidly to dynamics changes and domain shifts in the field.

\subsection{The Human--AI Alignment Gap (Deployment)}
This is often the most critical and overlooked gap. A mathematically optimal policy has limited value if it is not adopted by field staff or if it conflicts with unwritten social norms, political constraints, or \textbf{dynamic operational requirements}. Domain experts (e.g., frontline health workers and security personnel) hold essential tacit knowledge that is rarely captured in objectives or training data. The alignment gap arises when AI recommendations clash with human judgment or when static models fail to keep pace with emerging needs (e.g., major events or shifting priorities). Closing this gap requires mechanisms that incorporate evolving expert guidance directly into the policy-generation loop.

\paragraph{Public Safety and Operational Requirements.}
Our work pioneered computational game theory for operational security, beginning with the \textbf{ARMOR} deployment at \textbf{LAX in 2007}~\cite{pita2008deployed}. Despite its success (e.g., adoption by the \textbf{US Coast Guard} and \textbf{US Federal Air Marshals}, and \textbf{more than $\$100$ million in savings}), ARMOR exposed a key alignment challenge: agencies routinely faced \textbf{dynamic, hard-to-quantify operational requirements} (e.g., reallocations for global events such as the Olympics). Encoding these shifting priorities into optimized randomized schedules was cumbersome and error-prone, often requiring sustained manual intervention.

\paragraph{Health Infrastructure and Expert Alignment.} 
Similar challenges emerge in our work on optimizing healthcare infrastructure in Ethiopia, where the Ministry of Health is upgrading rural health posts to provide comprehensive services such as childbirth and postnatal care.
While algorithmic planning can maximize population coverage, scaling these solutions is often hindered by a persistent gap between quantitative optimization and the qualitative guidance provided by domain experts. 
To bridge this gap, we developed the Large language model and Extended Greedy (LEG) framework and the Health Access Resource Planner (HARP) tool~\cite{trabelsi2026}. 
Our LEG framework leverages LLMs to iteratively improve stakeholder alignment while maintaining strong coverage guarantees. 
Furthermore, our greedy and learning-augmented approaches in HARP ensure that resource allocation remains equitable across districts—satisfying persistent proportionality targets even under the uncertainty of online, multi-year budgets~\cite{choo2026optimizing}. 
Deployments in regions such as Afar and Somali demonstrate that the primary bottleneck in scaling is the need for a unified approach that synthesizes rigorous optimization with evolving operational constraints and human expertise.

\section{Generative AI for Solutions to the Deployment Bottleneck}

\subsection{Addressing the Alignment Gap with LLM Agents}

The human–AI alignment gap—the tension between mathematically optimal policies and real-world feasibility, which is especially acute in dynamic security and public health settings—stems largely from the difficulty of translating tacit human knowledge into formal algorithmic inputs. We argue that this gap can be addressed by systematically converting expert preferences, operational constraints, and contextual judgment into explicit, machine-readable objectives and constraints. \textbf{LLM Agents} are uniquely positioned to serve as the interface for this translation.

Concretely, LLM Agents can ingest natural-language guidance from domain experts (e.g., “Our priority this quarter is prioritizing health post upgrades in rural districts with high home-birth rates to improve access to maternal and postnatal care.,” or “Deploy 20\% more air marshals to the London route next week due to the Olympics”) and convert it into executable mathematical specifications—such as modified objective terms or hard constraints—for downstream optimization algorithms, including restless bandits, security games, and reinforcement learning–based planners \cite{behari2024dlm, kim2025preferencerobustness, Verma2025}.

\paragraph{LLM Agent Workflow.}
\begin{enumerate}
\item \textbf{User Guidance Acquisition:} The LLM Agent receives complex, unstructured inputs from field personnel or policymakers through text or voice, capturing high-level goals, operational restrictions, and contextual priorities.
\item \textbf{Constraint Formalization:} Using its reasoning and semantic parsing capabilities, the LLM Agent translates natural-language instructions—such as temporal, geographic, or policy-driven constraints—into formal mathematical constraints on the optimization problems or reward function changes (e.g., \texttt{max\_visits\_per\_week[village\_A] = 0} or \\ \texttt{reward = engaged + 3 * (lowest\_education or \\ second\_lowest\_education)}).
\item \textbf{Algorithmic Planning:} These formalized models are passed to off-the-shelf optimization solvers (e.g., game solvers or RMAB solvers), which compute resource allocation strategies that explicitly reflect real-world priorities and feasibility constraints. The LLM agent runs simulations and refines the reward functions or constraints until it satisfies the requirements \cite{behari2024dlm}.
\item \textbf{Human-in-the-Loop Solution Selection:} The user reviews and selects among a small set of candidate solutions generated under different interpretations or parameterizations, ensuring transparency, trust, and accountability \cite{kim2025navigating}.
\end{enumerate}

By embedding LLM Agents into the optimization loop, this approach equips domain experts with a mathematical decision-making toolkit that would otherwise require repeated, time-consuming interactions with algorithm designers. The resulting policies are not only near-optimal in a formal sense, but also aligned with on-the-ground constraints, interpretable by stakeholders, and substantially easier to adapt as operational conditions evolve.

\subsection{Addressing the Scarcity and Synthesis Gaps with Diffusion Models}

Diffusion models~\cite{ho2020denoising, song2021scorebased}, best known for high-fidelity generation in vision and language, are increasingly effective for modeling structured, high-dimensional objects that arise in AI4SI---including graphs, spatiotemporal dynamics, and combinatorial actions.
Their core strength is the ability to learn rich distributions from limited observations and to generate diverse samples that are statistically consistent with real-world structure.
This makes diffusion models a natural tool for addressing both the \textbf{observational scarcity gap}  and the \textbf{policy synthesis gap}.

\subsubsection{Generative Synthetic Social Networks}
For network-based interventions such as our HIV prevention efforts (Section~2.1), the lack of accurate and complete network data can be paralyzing: relational structure is partially observed, noisy, and often cannot be collected at scale due to logistical and operational constraints.
Diffusion models offer a principled way to mitigate this \textbf{observational scarcity gap} by learning a generative distribution over graphs from available network datasets and then sampling \emph{new} networks that are statistically consistent with the populations of interest.

In particular, diffusion-based graph generators can produce large collections of realistic \textbf{synthetic social networks} that preserve salient structural properties of real networks, such as degree heterogeneity, clustering, community structure, and homophily patterns.
This yields two practical benefits for AI4SI: (i) \emph{data amplification} for learning and evaluation, enabling downstream decision systems (e.g., influence maximization or adaptive testing policies) to be trained more reliably, and (ii) \emph{robust policy design} by training and validating decision rules over an ensemble of plausible network realizations rather than a single estimated graph, improving generalization when the true network is only partially observed.

\subsubsection{Scalable and Generalizable Policy Synthesis via Generation}
Diffusion models can help bridge the \textbf{policy synthesis gap} by enabling scalable policy representations and more robust learning under domain shift.
In AI4SI, they can model latent components of the environment---such as reward structure and transition dynamics---and can also directly parameterize stochastic policies.
Concretely, diffusion-based policy synthesis supports three complementary capabilities:

\begin{enumerate}
    \item \textbf{Scalable representations for combinatorial policies.}
When actions are combinatorial (e.g., choose 5 out of 50 regions), we cannot represent a full probability distribution over all subsets. Instead, we train a diffusion model to generate 'action blueprints'—continuous score vectors representing regional priorities. These continuous blueprints are deterministically projected into discrete, feasible actions via a constraint solver (e.g., a top-$k$ selection subject to routing limits). By leveraging the stochastic nature of the diffusion sampling process, we induce a complex distribution over the combinatorial action space, thereby circumventing the intractability of explicit enumeration."

    \item \textbf{Robust policy learning via generative environment modeling.}
    By learning distributions over rewards and transition dynamics, diffusion models expose policies to a family of plausible environments rather than a single point estimate~\cite{kong2025robust}.
    Training and evaluation under this generative ensemble can improve robustness and calibration under distribution shift and nonstationarity.

\item \textbf{Fast adaptation via transfer.}
Deployments often share structure (e.g., parks have similar terrain constraints; clinics have similar capacity limits), yet standard pipelines still require expensive retraining in every new site.
Diffusion models provide a principled way to transfer knowledge by \emph{starting generation from a source distribution that is already close to the target}, rather than from a generic base (e.g., Gaussian noise)~\cite{de2021diffusion, lipman2023flow}.
This reuse of prior information can substantially reduce the amount of new interaction data needed before the policy performs well~\cite{kong2025composite, kong2025generative}.

\end{enumerate}

\section{Conclusion}

The future of AI for Social Impact hinges on our ability to move beyond local algorithmic success and achieve scalable, real-world deployment. The deployment bottleneck---defined by the observational scarcity, policy synthesis, and human-AI alignment gaps---is the principal inhibitor of this scale.

We argue that generative AI—particularly LLM agents and diffusion models—offers a practical path to overcoming this systemic deployment bottleneck. LLM agents integrate the vital, tacit knowledge of human experts and dynamic operational needs into the policy loop, solving the alignment gap exposed by our long-term security and health deployments. Diffusion models generate realistic synthetic data and support faster cross-domain policy generation, helping to solve both the data scarcity and policy synthesis gaps.

By embedding these generative technologies into the AI4SI lifecycle, we can move towards a future of truly scalable, human-aligned, and trustworthy AI that consistently optimizes resources and drives measurable positive change for communities across the globe. This represents a critical step toward maximizing the societal promise of artificial intelligence.

\bibliography{refs}
\bibliographystyle{alpha}

\end{document}